\documentclass[intlimits,twoside,a4paper]{article}

\usepackage{amsmath,amssymb}

\usepackage[T2A]{fontenc}
\usepackage[cp1251]{inputenc}

\usepackage[eqsecnum]{cmpj2}

\usepackage{mathrsfs}
\usepackage{bbold}
\usepackage{float}
\usepackage{graphicx}
\usepackage{bm}
\usepackage{booktabs}
\usepackage{soul}
\usepackage{pstricks}

\RequirePackage{color}
\definecolor{MyDarkGreen}{rgb}{0.02,0.60,0.06}

\def\bsigma{{\pmb{\sigma}}}

\def\W{{\pmb{\cal W}}}

\def\nbOne{\pmb{\mathbb 1}}

\def\be{\begin{equation}}
\def\ee{\end{equation}}
\def\bey{\begin{eqnarray}}
\def\eey{\end{eqnarray}}

\def\mats{{\vac s}}

\def\vac#1{{\bf{#1}}}
\def\I{\ \!\vec{\rm \i}}
\def\J{\ \!\vec{\rm \j}}

\issue{2017}{20}{1}{13702}
\doinumber{10.5488/CMP.20.13702}

\title[Rashba spin-orbit interaction enhanced by graphene in-plane deformations]{Rashba spin-orbit interaction enhanced by graphene in-plane deformations\thanks{A paper dedicated to our friend Yurij Holovatch on the occasion of his 60th birthday.}}
\author[B. Berche, F. Mireles, E. Medina]{B. Berche\refaddr{label1,label2}, F. Mireles\refaddr{label3}, E. Medina\refaddr{label1,label2,label4}}
\addresses{
\addr{label1} Groupe de Physique Statistique, Institut Jean Lamour, Universit\'e de Lorraine, \\54506 Vandoeuvre-les-Nancy, France
\addr{label2}Centro de F\'isica, Instituto Venezolano de Investigaciones Cient\'ificas, 21827, Caracas, 1020 A, Venezuela
\addr{label3} Centro de Nanociencias y Nanotecnolog\'ia, Universidad Nacional Aut\'onoma de M\'exico, \\ Apdo. Postal 14, 22800 Ensenada B.C., M\'exico
\addr{label4} Yachay Tech, School of Physical Sciences \& Nanotechnology, 100119-Urcuqu\'i, Ecuador
}

\authorcopyright{B. Berche, F. Mireles, E. Medina, 2017}
\date{Received January 2, 2017, in final form February 8, 2017}

\begin{document}
\maketitle

\begin{abstract}
Graphene consists in a single-layer carbon crystal where 2$p_z$ electrons display a linear dispersion relation in the vicinity of the Fermi level, conveniently described by a massless Dirac equation in $2+1$ spacetime. Spin-orbit effects  open a gap in the band structure and offer perspectives for the manipulation of the conducting electrons spin.
Ways to manipulate spin-orbit couplings in graphene have been generally assessed by proximity effects to metals
that do not compromise the mobility of the unperturbed system and are likely to induce strain in the graphene layer. In this work we explore the $\rm{U(1)}\times SU(2)$ gauge fields
that result from the uniform stretching of a graphene sheet under a perpendicular electric field. Considering such deformations
is particularly relevant due to the counter-intuitive enhancement of the Rashba coupling between 30-50\% for small bond
deformations well known from tight-binding and DFT calculations. We report the accessible changes that can be operated in the
band structure in the vicinity of the K points as a function of the deformation strength and direction.
\keywords{graphene, spin-current, spin-orbit interaction}
\pacs 72.80.Vp, 75.70.Tj, 11.15.-q
\end{abstract}

\section{Introduction}\label{sec1}

Modifying the interactions present or lacking in graphene has been a topic of continued interest since
perturbations by, e.g., proximity effects \cite{Peralta16}, can generate gaps \cite{Zhou,Pastawski} for semiconducting properties,
spin splitting for spintronic effects \cite{KaneMele,Wang07}, spin-orbit interactions for topologically protected edge currents \cite{Fertig},
spin alignment to induce anomalous Hall effects \cite{Wang} and RKKY interactions \cite{Macdonald}. Regarding
spin-orbit (SO) interactions, the intrinsic contribution in graphene depends
directly on the atomic SO coupling of carbon, ($\sim 6$~meV). Nevertheless, it is a second order
effect in flat graphene with energies in the range of {\textmu}eV. To enhance such an
interaction one can resort to the bending of the graphene sheet \cite{MirelesEtAl}, as it occurs in nanotubes \cite{AndoSO}. This results
in a SO interaction increase of three orders of magnitude due to a change in the $p_z$ orbital overlaps.
On the other hand, bringing the graphene in close proximity to a gold interface can increase the Rashba
coupling \cite{Marchenko} both due to the interface electric field caused by charge transfer and due to the
strong atomic SO interaction of gold. Depending on the registry of the two materials, the SO can increase to $\sim 70$~meV.

Here, we explore another mechanism for SO enhancement, making use of uniform lattice
deformations. We take advantage of the non-intuitive behaviour of the Rashba coupling in planar
graphene that depends inversely proportional to the $\sigma$ bond length \cite{Huertas,Konschuh}.
The stretching of the bonds lead to an increased SO coupling when an electric field, perpendicular
to the graphene plane, is present. In this work we will first derive a uniform deformation
field that will stretch and rotate the base graphene lattice. We will then derive the Hamiltonian corresponding
to such deformations which can
be mapped onto an equivalent  U(1) and SU(2) gauge field theory \cite{JinLiZhang06,MedinaLopezBerche08,Tokatly08,BercheChatelainMedina,BercheMedinaEJP}, once an expansion
around the K point is performed. We then derive the energy spectrum of the problem as a function
of the stretching intensity and the rotation angle. We find that a uniform strain can control both the formation
of a substantial gap for strains less than 15\% and modulate the chiral spin splitting of the band structure. The
strain angle also controls the spin splitting, which becomes less sensitive to strain as the strain angle increases. Let us also mention a recent study that reported a spin filter/valley filter via strain induced Rashba SO interaction and magnetic barrier  \cite{WuEtAl2016}.

\section{Undeformed Hamiltonian in the vicinity of the K point}
\begin{figure}[!b]
\centering
\includegraphics[width=6cm]{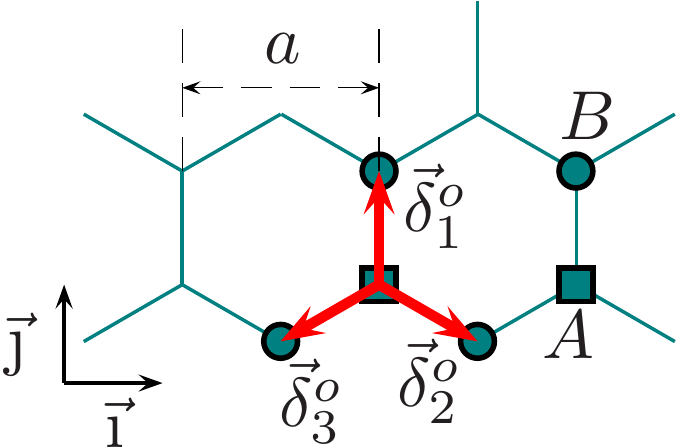}
\caption{(Color online) The primitive cell in unstrained graphene (pair of square and circle sites) and the vectors $\vec\delta_i^o$
joining the A site to its nearest neighbours. The lattice parameter $a$ indicates
a reference length scale of the lattice.}
\label{figNotations}
\end{figure}

We are interested in the modification of graphene electronic properties due to sheet deformations
which lead to changes in the hopping parameters within a tight-binding (TB) approach \cite{PastawskiMedina,VarelaEtAl16}.
This problem was elegantly addressed in the U(1) gauge theory context \cite{Vozmediano}. Here, we
extend this analysis to include spin-orbit (SO) effects which can be enhanced due to in-plane deformations.
We use the convention that one of the C$-$C bonds is chosen along the $y$ direction of the lattice in
such a way that the nearest neighbour lattice vectors, between A and B sublattices, are denoted as
$\vec\delta_1^o=(a/\sqrt{3})\J $, $\vec\delta_2^o=(a/2)\I -(a/2\sqrt 3)\J $ and $\vec\delta_3^o=-(a/2)\I -(a/2\sqrt 3)\J $. The modulus
of the nearest neighbour vectors is given by the C$-$C distance,
$|\vec\delta_i^o|=a/\sqrt 3= 1.42$~{\AA} where $a=2.46$~{\AA} is the unit cell basis vector's modulus (see figure~\ref{figNotations}).

In  the reciprocal space, the Brillouin zone (BZ) is hexagonal, with Dirac points at the edges, where the dispersion relation is linear. The coordinates $\vec K_\zeta$ of the two inequivalent Dirac points are labelled by a (valley) parameter $\zeta=\pm 1$, $\vec K_\zeta=\zeta(4\pi/3a,0)$, and the vicinity of these Dirac points in momentum space is parameterized by the wave vectors
$\vec k=\vec K_\zeta+\vec p/\hbar$, hence $k_x=\zeta (4\pi/3a)+p_x/\hbar$, $k_y=p_y/\hbar$.

In the continuum limit, the Hamiltonian follows from the expansion of the TB matrix elements in powers of $\vec p/\hbar$. If one only retains the
kinetic energy (KE), the non-diagonal matrix elements describing the nearest neighbour (nn) hopping are given by
\be
(H)_{\text{AB}}=\sum_i (-t) \re^{\ri\vec k\cdot\vec\delta_i^o}=\frac{\sqrt 3at}{2\hbar}(p_x-\ri p_y)+O(|\vec p|^2/\hbar^2),
\label{TBKE}
\ee
where the sum is over nn sites and the prefactor proportional to the hopping amplitude defines the Fermi velocity $v_{\text F}=\sqrt{3}at/2\hbar$ with $t=V_{pp\pi}$, the hopping integral between $p_z$ clouds in graphene. The other non-diagonal matrix elements follow from complex conjugation, $(H)_{\text{BA}}=(H)_{\text{AB}}^*$, and the diagonal elements $(H)_{\text{AA}}=(H)_{\text{BB}}$ are chosen to be zero fixing the Fermi
energy. The primitive cell Hamiltonian can be collected into the matrix expression (from now on, we will focus on the $\zeta=+1$ valley without loss of generality)
\be
\vac H_0=v_{\text F}(\bsigma_x p_x+\bsigma_y p_y),\label{ContinuumKE}
\ee
where dimensionless Pauli matrices $\bsigma_i$ have been introduced to describe a pseudo-spin degree of freedom. In what follows, bold fonts
are used to denote matrices in pseudo-spin and/or spin space.

When a graphene sheet is deformed, the lattice is subjected to local modifications that in general change the nearest neighbour vectors $\vec\delta_i^o$ and thus change the  continuum Hamiltonian in equation~(\ref{ContinuumKE}). Since the hopping amplitudes $t$ are
themselves functions of the nearest neighbour distance (typically exponentially decaying functions of the distance), it is simpler and more transparent to model an arbitrary sheet deformation by a change in the hopping integrals $t\to t_i$ \cite{Vozmediano,CastroNetoArxiv}. Here, we disregard bond angle effects on hopping
integrals \cite{AndoPhonons} which are important for electron-phonon coupling.

The effect of deformations on the kinetic energy is well known, but its role in the spin-orbit interaction (SOI) have not been fully addressed. Here,
we focus on the Rashba spin-orbit (RSO) interaction which is due to both
the atomic SOI of graphene, and an externally applied electric field perpendicular to the graphene surface. The latter can be due to
either a charge transfer to a nearby substrate or an applied gate voltage. For a perfect lattice, non-diagonal RSO amplitudes are given, in the TB approach, by terms of the form \cite{KaneMele2}
\begin{align}
(H_{\text R})_{\text{AB}}&=-\ri\frac{2}{\sqrt{3}a}\sum_i \lambda_{{\text R}_i}\re^{\ri\vec k\cdot\vec\delta_i}\vec\delta_i\cdot(-\mats_y\I  +\mats_x\J )
\nonumber\\
&=\lambda_{\text R}\left\{
-\ri\mats_x-\mats_y+\frac{1}{2\sqrt 3} \left[ \frac{p_x a}{\hbar}(-\ri\mats_x+\mats_y) +\frac{p_ya}{\hbar}(\mats_x+\ri\mats_y) \right]
\right\}+O(|\vec p|^2/\hbar^2).
\label{TBRSO}
\end{align}
Here, assuming unstrained  graphene $\lambda_{{\text R}_i}\equiv\lambda_{\text R}$ is a constant,  having the dimensions  of energy and ranges between 13 to 225~meV depending upon the substrate \cite{Varykhalov,Dedkov,Rader,Marchenko}.
$\mats_i$ are the dimensionless spin Pauli matrices.
Taking into account the kinetic energy corrections due to the SOI,  the Rashba Hamiltonian
is then
\be
\vac H_{\text R}=\lambda_{\text R}
\left\{\bsigma_y\mats_x-\bsigma_x\mats_y
{
+\frac{1}{2\sqrt{3}}
 \left[ \frac{p_x a}{\hbar}(\bsigma_y\mats_x+\bsigma_x\mats_y) +\frac{p_ya}{\hbar}(\bsigma_x\mats_x-\bsigma_y\mats_y) \right]
}
\right\}
\label{ContinuumRSO}
\ee
in the vicinity of the K point. Contrary to equation~(\ref{ContinuumKE}),  this expression has a $4\times 4$ matrix structure, where
$\bsigma_i\mats_j$ is a short notation for $\bsigma_i\otimes\mats_j$, hence, when compared to SOI, the purely kinetic term is implicitly multiplied by the identity matrix in spin space  $\nbOne_s$.

\section{Gauge fields in the deformed graphene sheet Hamiltonian}\label{sec2}
\begin{figure}[!t]
\begin{center}
\includegraphics[width=9cm]{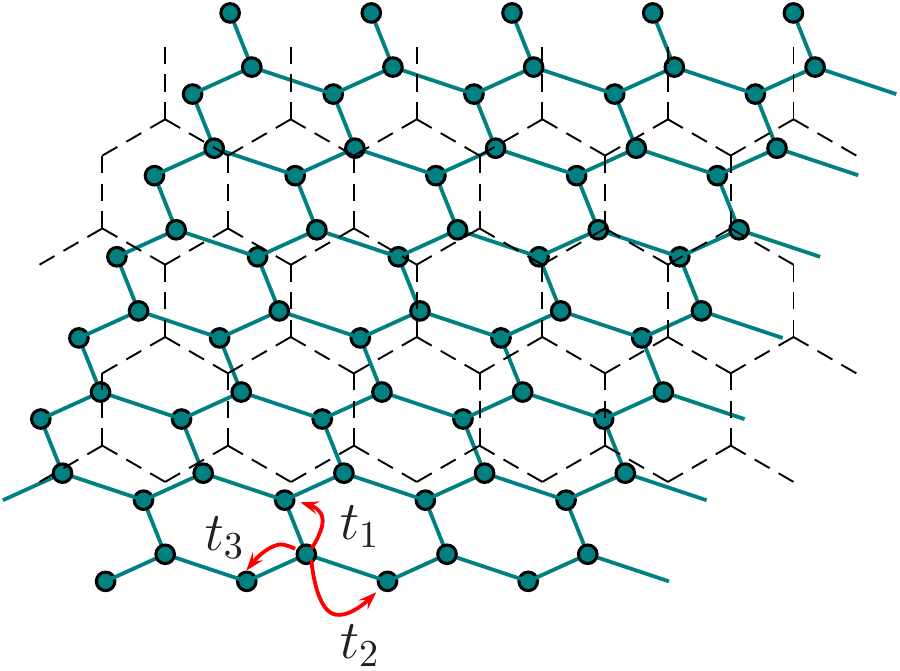}
\vspace{0cm}
\caption{(Color online) Homogeneously deformed graphene sheet, parameterized by strain $\varepsilon$ and strain angle $\theta$ as described in equations~(\ref{deformations})--(\ref{eqdeltat}).}
\label{figDeformedSheet}
\end{center}
\end{figure}

Let us now consider that the graphene sheet is subjected to a uniform tension in the plane resulting in a tensile strain in a given direction, in the coordinate system $x^{\prime}$-$y^{\prime}$, oriented at an angle $\theta$ relative to the lattice coordinate system  $x$-$y$.  Such strain would  induce a uniform deformation of the lattice (figure~\ref{figDeformedSheet}). The deformation is characterized by the strain tensor $\bm\epsilon^{\prime}$ in the $x^{\prime}$-$y^{\prime}$ system. For a graphene sheet, the only nonzero deformations are $\epsilon^{\prime}_{xx}$ and $\epsilon^{\prime}_{yy}$, so
\begin{equation}
\bm\epsilon^{\prime} = \left(
\begin{array}{cc}
 \epsilon^{\prime}_{xx} &0\\
 0 & \epsilon^{\prime}_{yy} \\
\end{array}
\right),
\end{equation}
\noindent where the uniaxial strain components are related to each other through the Poisson ratio $\sigma$  by $\epsilon^{\prime}_{yy}=-\sigma\epsilon^{\prime}_{xx}$, with $\sigma=0.165$ in graphite \cite{Poisson_ratio, Blakslee,Pereira-Neto}. The strain tensor in the lattice coordinate system is given by $\bm\epsilon=U^{\dagger}\bm\epsilon^{\prime}U$, where the matrix elements of the rotation matrix $U$ are $U_{\mu\nu}=  \delta_{\mu\nu}\cos\theta +(-1)^{\nu}\sin\theta (1-\delta_{\mu\nu}) $, with $\{\mu,\nu\}=1,2.$ The deformation induces a change in the nearest-neighbor lattice vectors $\vec\delta_i^o$ as follows, $\vec \delta_i=(\nbOne+\bm\epsilon) \vec \delta_i^o$, with $i=1,2,3$,  leading to (in units of $a_o=a/\sqrt{3}$, i.e., all $\delta_i$'s to be understood as $\delta_i/a_o$),
\begin{equation}
|\vec \delta_i|\simeq 1+\epsilon_{11}(\delta_{x i}^o)^2+\epsilon_{22}(\delta_{y i}^o)^2
+2\epsilon_{12}\delta_{x i}^o\delta_{y i}^o\,,
\end{equation}
where the strains are defined by
\begin{align}
\label{deformations}
\epsilon_{11} & =  \epsilon({\cos^2\theta}-\sigma \sin^2\theta), \\
\epsilon_{12} & =  \epsilon(1+\sigma){\cos\theta}\sin\theta, \\
\epsilon_{22} & = \epsilon( {\sin^2\theta}-\sigma \cos^2\theta).
\end{align}
We use as a tunable parameter the uniaxial strain $\varepsilon=\epsilon_{xx}$.
The deformations lead to a modification of the hopping amplitudes which decay exponentially with the nn atomic distance. For the
case of the kinetic term one has
\be
t_i\equiv t_{pp\pi}(|\vec \delta_i|)=t\re^{-\beta(|\vec\delta_i|/a_o-1)},\label{3.6}
\ee
where $\beta\simeq 3.37$ \cite{Pereira-Neto,CastroNetoGuinea07}.
Starting from the TB expressions and allowing only for nn hopping
amplitudes $t_i$, with $i=1,2,3$ according to the labeling of the n.n. vectors $\vec \delta_i$, equation~(\ref{TBKE}) for the kinetic energy
becomes
\bey
(H')_{\text{AB}}=-t\sum_i  \re^{\ri\vec k\cdot\vec\delta_i}+\sum_i (t-t_i) \re^{\ri\vec k\cdot\vec\delta_i}=v_{\text F}(p_x-\ri p_y-{\cal A}_x+\ri{\cal A}_y)
+ \hbox{ second order terms},\label{TBKEdeformed}
\eey
with
\begin{align}
{\cal A}_x&=\frac 2{\sqrt 3}\frac {\hbar}{at}\left[t_1-\frac 12(t_2+t_3)\right]\equiv\frac{2\hbar} {\sqrt 3 a}\frac{\delta t'}{t}\,,\label{eqdeltatprime}\\
{\cal A}_y&=\frac {\hbar}{at}(t_3-t_2)\equiv\frac\hbar a\frac{\delta t}{t}\,.
\label{eqdeltat}
\end{align}
Both terms above vanish in the case of the undeformed graphene sheet. The notation suggests the introduction of an Abelian gauge
field \cite{Vozmediano} $\vec {\cal A}={\cal A}_x\I+{\cal A}_y\J$ (here, in two space dimensions)
in order to describe the effect of sheet deformations. The second order terms which are neglected in
equation~(\ref{TBKEdeformed}) comprise $O(|\vec p|^2a^2/\hbar^2)$ terms as well as products of
order $O(|\vec p|a\delta t/\hbar t)$ where $\delta t/t$ represents any dimensionless combination involving
the hopping amplitudes and which vanish for the undeformed graphene sheet, e.g., of the form $\delta t=(t_3-t_2)$ or $\delta t'=t_1-\frac 12(t_2+t_3)$ in equations~(\ref{eqdeltatprime})--(\ref{eqdeltat}).
In this approximation, the continuum limit counterpart of (\ref{TBKEdeformed}) is now
\be
\vac H_0=v_{\text F}\big[\bsigma_x (p_x-{\cal A}_x)\nbOne_s+\bsigma_y (p_y-{\cal A}_y)\nbOne_s\big].\label{ContinuumKEdeformed}
\ee

The Rashba term also needs to be corrected to account for space dependent hopping amplitudes. However, the origin
of the coupling between nearest neighbours is quite different, and arises via the atomic SOI and the Stark interaction (due to the external electric field). Within tight-binding approach, the Rashba parameter strength for the unstrained case is given to leading order by \cite{Huertas,Konschuh}
\begin{equation}
\lambda_{\text R}^{(0)}=\frac{2eE z_{sp}}{3V_{sp\sigma}}\xi_p\,,
\end{equation}
where $E$ is the external electric field, $eE z_{sp}$ is the matrix element of the Stark Hamiltonian between $s$ and $p_z$
orbitals, $\xi_p$ is the matrix element of the atomic SOI (carbon) coupling the $p_z$ orbitals and the bonded in plane $p_x$, $p_y$ orbitals.
Finally, $V_{sp\sigma}$ is the matrix element coupling the $s$ orbital in one carbon to the $p_x$, $p_y$ orbital of the n.n. atom. It is the latter bond
that is stretched by the deformation field. According to DFT and tight-binding calculations \cite{Konschuh}, the Rashba parameter   shows an exponential increase due to the lattice constant stretching. It is controlled by the decay of the matrix element $V_{sp\sigma}$ through the increase of interatomic carbon-carbon distance.

\begin{figure}[!b]
\begin{center}
\includegraphics[width=0.5\linewidth]{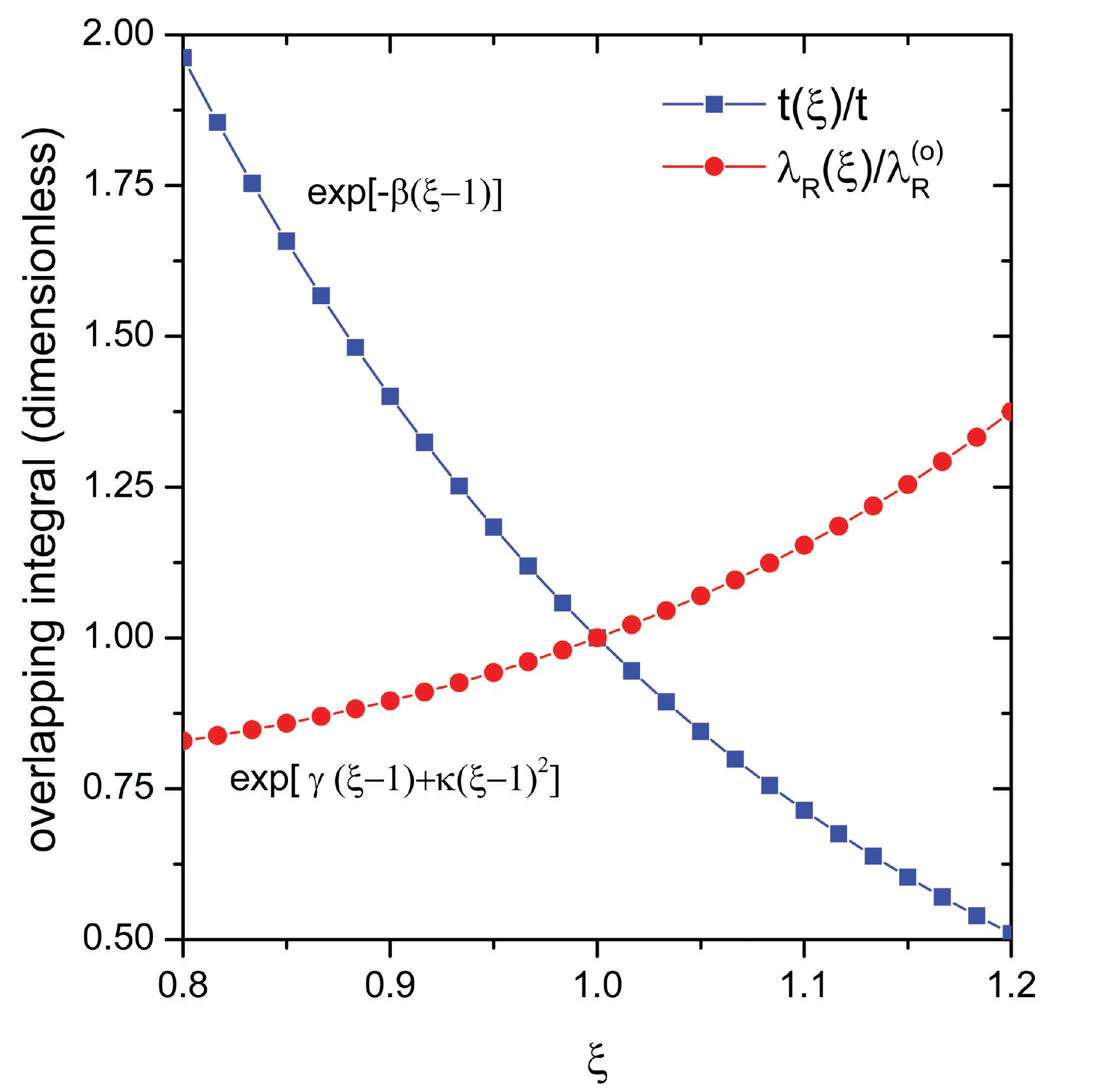}
\caption{(Color online) Exponential behavior of the hopping parameter $ t(\xi)/t$ [equation~(\ref{3.6})] and Rashba coupling strength $\lambda_{\text R}(\xi)/\lambda_{\text R}^{(0)}$ [equation~(\ref{3.12})] and as a function of  a given stretched C--C lattice constant $\xi=|\vec\delta_i|/a_0$. Here $\beta=3.37$, $\gamma=1.265$, and $\kappa=1.642$. }
\label{fig3}
\end{center}
\end{figure}

We can express $\lambda_{{\text R}_{i}}=\lambda_{\text R}^{(0)}[V_{sp\sigma}^{(0)}/V_{sp\sigma}]=\lambda_{\text R}^{(0)} \tau^{sp\sigma}_i$ for each bond, where $V_{sp\sigma}$  is the stretched hopping matrix element. By fitting the numerical values of the dependence on the lattice constant arising from the hopping parameter $V_{sp\sigma}$ reported in \cite{Konschuh}, we can thus model $\tau^{sp\sigma}_i$ as a dimensionless ratio given by
\be
\tau^{sp\sigma}(|\vec \delta_i|)= \!\re^{\gamma(|\vec\delta_i|/a_0-1)}\re^{\kappa (|\vec\delta_i|/a_0-1)^2 },
\label{3.12}
\ee
where $\gamma=1.265$ and $\kappa=1.642$. For illustration, in figure~\ref{fig3} we plot the dependence of  the hopping parameter $ t(\xi)/t$ and Rashba coupling strength $\lambda_{\text R}(\xi)/\lambda_{\text R}^{(0)}$ as a function of  a given stretched lattice constant $\xi=|\vec\delta_i|/a_0$.

Henceforth, to zeroth order in $p_x$, $p_y$, equation~(\ref{TBRSO}) obeys the expression
\begin{align}
(H'_{\text R})_{\text{AB}}&=-\ri\lambda_{\text R}^{(0)}\frac{2}{\sqrt{3}a}\sum_i \tau_i^{sp\sigma} \re^{\ri\vec k\cdot\vec\delta_i}\vec\delta_i\cdot(-\mats_y\I  +\mats_x\J )
=\frac{\lambda_{\text R}^{(0)}}{6}\Big[
\left(4\tau_1^{sp\sigma}+\tau_2^{sp\sigma}+\tau_3^{sp\sigma}\right)(-\ri\mats_x)\nonumber\\
&\quad-\sqrt{3}\left(\tau_2^{sp\sigma}-\tau_3^{sp\sigma}\right)\mats_x+\sqrt{3}
\left(\tau_2^{sp\sigma}-\tau_3^{sp\sigma}\right)(-\ri\mats_y)
-3\left(\tau_2^{sp\sigma}+\tau_3^{sp\sigma}\right)\mats_y\Big],
\end{align}
thus, the full Rashba Hamiltonian, to lowest order is expressed as
\bey
\vac H_{\text R}^{(0)}=
\frac{\lambda_{\text R}^{(0)}}{6}\left[
\left(4\tau_1^{sp\sigma}+\tau_2^{sp\sigma}+\tau_3^{sp\sigma}\right)\bsigma_y\mats_x
-3\left(\tau_2^{sp\sigma}+\tau_3^{sp\sigma}\right)\bsigma_x\mats_y
-\sqrt{3}\left(\tau_2^{sp\sigma}-\tau_3^{sp\sigma}\right)(\bsigma_x\mats_x-\bsigma_y\mats_y)
\right].
\label{ContinuumRSOdeformed}
\eey
It is interesting to note that together with an asymmetric Rashba type of interaction, there also appears a regular Dresselhaus spin-orbit (DSO) interaction $\propto(\bsigma_y\mats_y-\bsigma_x\mats_x)$ when the hopping parameters $\tau_2^{sp\sigma}$ and $\tau_3^{sp\sigma}$ are unequal, no matter the
parameters of the deformation.

As we have done in the case of the KE, we can introduce here {\em non-Abelian} gauge fields \cite{YangMills54} to describe the effect of deformation on the SOI, namely write equation~(\ref{ContinuumRSOdeformed}) as
\be
\vac H_{\text R}^{(0)}=-v_{\text F}\left(\bsigma_x{\cal W}^1_x\mats_x+\bsigma_x{\cal W}^2_x\mats_y
+\bsigma_y{\cal W}^1_y\mats_x+\bsigma_y{\cal W}^2_y\mats_y\right)
\ee
with the components ${\cal W}_i^\beta$ of the non-Abelian gauge field carrying an ordinary space index
$i=1,2$ and an internal superscript $\beta=1,2,3$ which specifies the spin component to which the term couples. Here, one has
\begin{align}
{\cal W}_x^1&=-\frac{\sqrt{3}\lambda_{\text R}^{(0)}}{6v_{\text F}}\left(\tau_2^{sp\sigma}-\tau_3^{sp\sigma}\right),\qquad
{\cal W}_y^1=\frac{\lambda_{\text R}^{(0)}}{6v_{\text F}}\left(4\tau_1^{sp\sigma}+\tau_2^{sp\sigma}+\tau_3^{sp\sigma}\right),\\
{\cal W}_x^2&=-\frac{\lambda_{\text R}^{(0)}}{2v_{\text F}}\left(\tau_2^{sp\sigma}+\tau_3^{sp\sigma}\right),\qquad
{\cal W}_y^2=\frac{\sqrt{3}\lambda_{\text R}^{(0)}}{6v_{\text F}}\left(\tau_2^{sp\sigma}-\tau_3^{sp\sigma}\right),
\end{align}
and there is no coupling to the third component $\mats_z$.
The non-Abelian character can be made explicit if one calculates the commutator between
$\W_i\equiv{\cal W}_i^\alpha\mats_\alpha$ (where  contraction over the superscript $\alpha=1,2$ is understood) and
$\W_j\equiv{\cal W}_j^\beta\mats_\beta$, i.e.
$[\W_i^\alpha,\W_j^\beta]=\ri\hbar\epsilon_{\alpha\beta\rho}{\cal W}_i^\alpha{\cal W}_j^\beta\mats_\rho$, with $\epsilon_{\alpha\beta\rho}$ the
fully antisymmetric tensor.
The non-Abelian gauge field components have the dimensions of momentum.
Let us define  $\delta \tau^{sp\sigma}=\tau_2^{sp\sigma}-\tau_3^{sp\sigma}$. Then, in a rather weak strain case ($\varepsilon<0.05$), the expression $(4\tau_1^{sp\sigma}+\tau_2^{sp\sigma}+\tau_3^{sp\sigma})/6\simeq(6+6\delta\tau^{sp\sigma})/6\simeq 1$ and therefore, ${\cal W}_x^1= -{\cal W}_y^2=\lambda_{\text R}^{(0)}\delta \tau^{sp\sigma}/(2\sqrt 3v_{\text F})$,  and ${\cal W}_y^1=-{\cal W}_x^2\simeq -\lambda_{\text R}^{(0)}/v_{\text F}$.

Altogether, the Hamiltonian that comprises the modified kinetic energy and the modified Rashba spin-orbit interaction may be written in terms of gauge fields as
\bey
\vac H=
v_{\text F}\left[\bsigma_x (p_x\nbOne_s-{\cal A}_x\nbOne_s-\W_x\cdot \mats )+\bsigma_y (p_y\nbOne_s-{\cal A}_y\nbOne_s -\W_y\cdot \mats
)\right].
\label{ContinuumTotdeformed}
\eey
We note that the intrinsic spin-orbit contribution $\propto\bsigma_z\mats_z$ which, being associated to next-nearest-neigh\-bour hopping terms,
is an order of magnitude lower so it  can safely be neglected.
Note that in contrast to \cite{WuEtAl2016,CastroNetoArxiv}  we have a non-Abelian
gauge field arising from the stretching of the lattice. In this work, stretching
produces both changes in the velocity and in the spin couplings.
We explicitly incorporate the change of the Rashba-parameter due to the lattice deformation,
which is controlled by the decay of the  hopping $V_{sp\sigma}$ with the nn interatomic distance,
while in \cite{WuEtAl2016} the Rashba parameter is just a constant in the strained region.

In conclusion, we have now a Hamiltonian to dominant order in $\delta t/t$ and $\delta \tau^{sp\sigma}$, changes that
can be cast as an effective U(1) and as a SU(2) gauge potential, respectively. We will now exactly diagonalize this Hamiltonian
in the previous approximation to see the effects on the band structure.

\section{Effect of the sheet deformation on the band structure}\label{sec3}

The energy spectrum in the continuum limit for graphene with uniaxial deformations and Rashba spin-orbit interaction is described by  the eigenvalues of the Hamiltonian (\ref{ContinuumTotdeformed}). We first notice that due to the small factor $\delta \tau^{sp\sigma}$,  the terms ${\cal W}_x^1$  and ${\cal W}_y^2$ are typically two orders of magnitude smaller than ${\cal W}_y^1$  and ${\cal W}_x^2$, then the dispersion energies are explicitly given by

\begin{equation}
E_{\eta}^{(s)}=\eta \,v_{\text F}\left\{\lambda_{\text{R}_1}^2 +\lambda_{\text{R}_2}^2 + (p_x-{\cal A}_x)^2+(p_y-{\cal A}_y)^2+2s \left[\lambda_{\text{R}_1}^2 \lambda_{\text{R}_2}^2 + \lambda_{\text{R}_1}^2 (p_x-{\cal A}_x)^2+\lambda_{\text{R}_2}^2 (p_y-{\cal A}_y)^2\right]^{\frac{1}{2}}  \right\}^{\frac{1}{2}}
\end{equation}

\begin{figure}[!b]
\begin{center}
\vspace{-0.8cm}
\includegraphics[width=11cm]{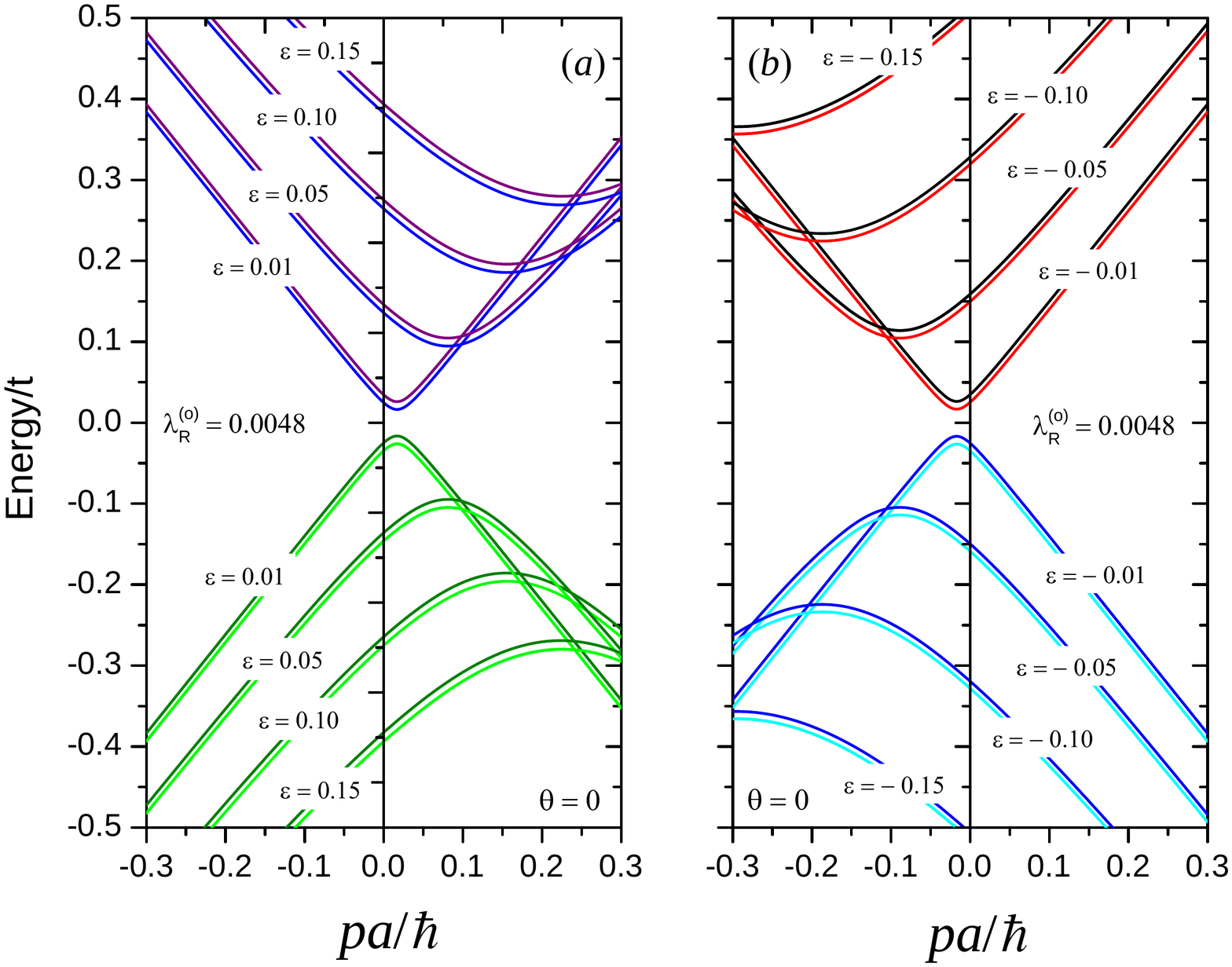}
\vspace{-0.1cm}
\caption{(Color online) Band structure for a graphene sheet under a simultaneous presence of relatively weak Rashba spin-orbit coupling [$\lambda_{\text R}^{(0)}=0.0048$ in  units of $t$] and different values of uniaxial strains in the plane. The band pairs correspond to the chiral spin splitting due
to the SOI. A large gap, $\sim0.5 t$, in the spectrum is produced by strains smaller than 15\%.}
\label{fig4}
\includegraphics[width=11cm]{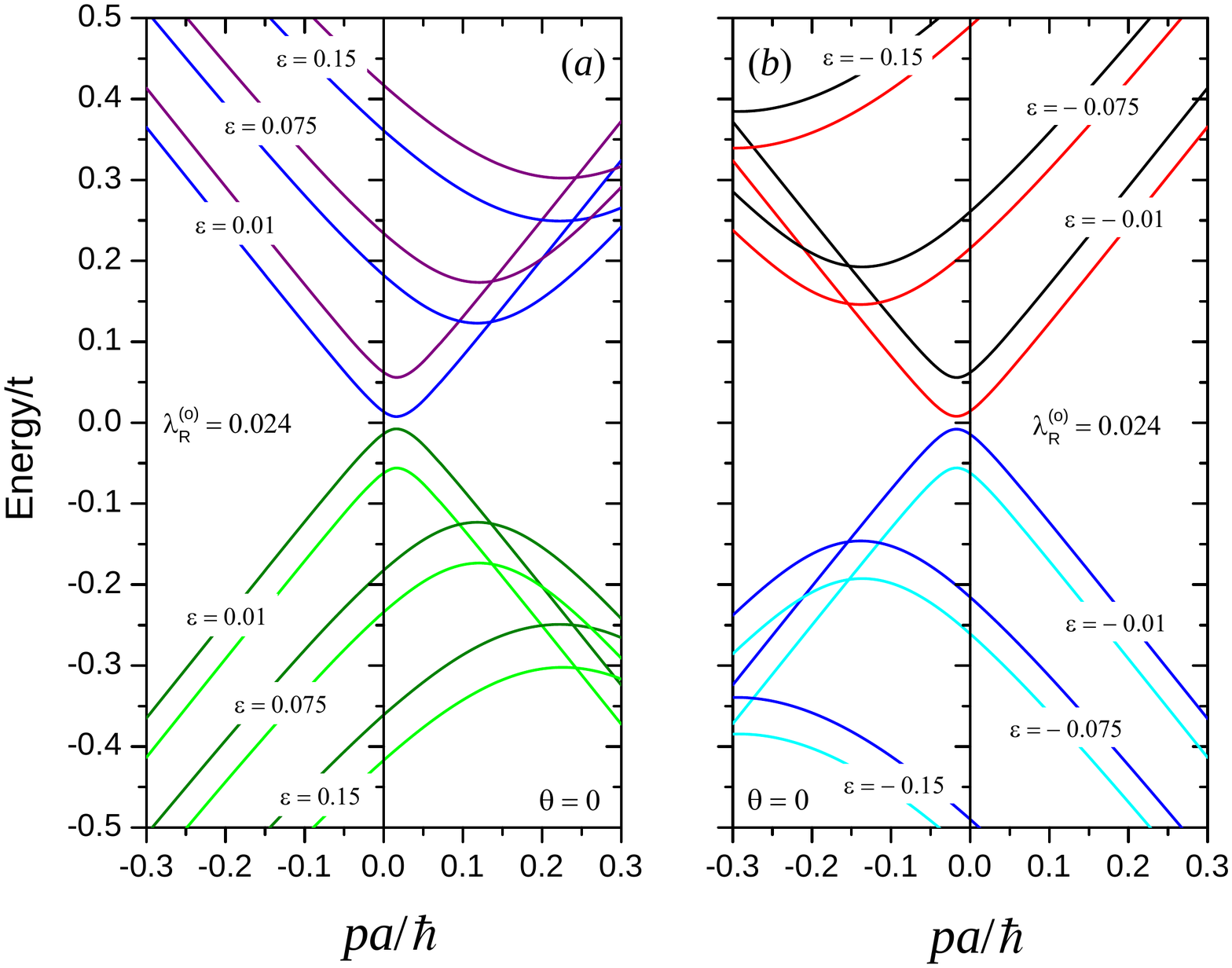}
\vspace{-0.1cm}
\caption{(Color online) Band structure for a graphene sheet under a simultaneous presence of  Rashba spin-orbit coupling [$\lambda_{\text R}^{(0)}=0.024$ in  units of $t$] and different values of uniaxial strains in the plane.  }
\label{fig5}
\end{center}
\end{figure}

\noindent with the definitions $\lambda_{\text{R}_1}={\cal W}_y^1$ and $\lambda_{\text{R}_2}={\cal W}_x^2$. The labels $\eta=+/-$ and $s=+/-$ denote the electron/hole  and spin chirality, respectively.
In the absence of Rashba spin-orbit coupling [$\lambda_{\text R}^{(0)} =0$], the band spectrum simplifies, as expected from the gauge approach solely applied to the kinetic energy, to
\begin{equation}
E_{\eta}^{(s)}=\eta v_{\text F}\left[(p_x-{\cal A}_x)^2+(p_y-{\cal A}_y)^2\right]^{\frac{1}{2}}.
\end{equation}

\begin{figure}[!t]
\begin{center}
\vspace{-0.3cm}
\includegraphics[width=11cm]{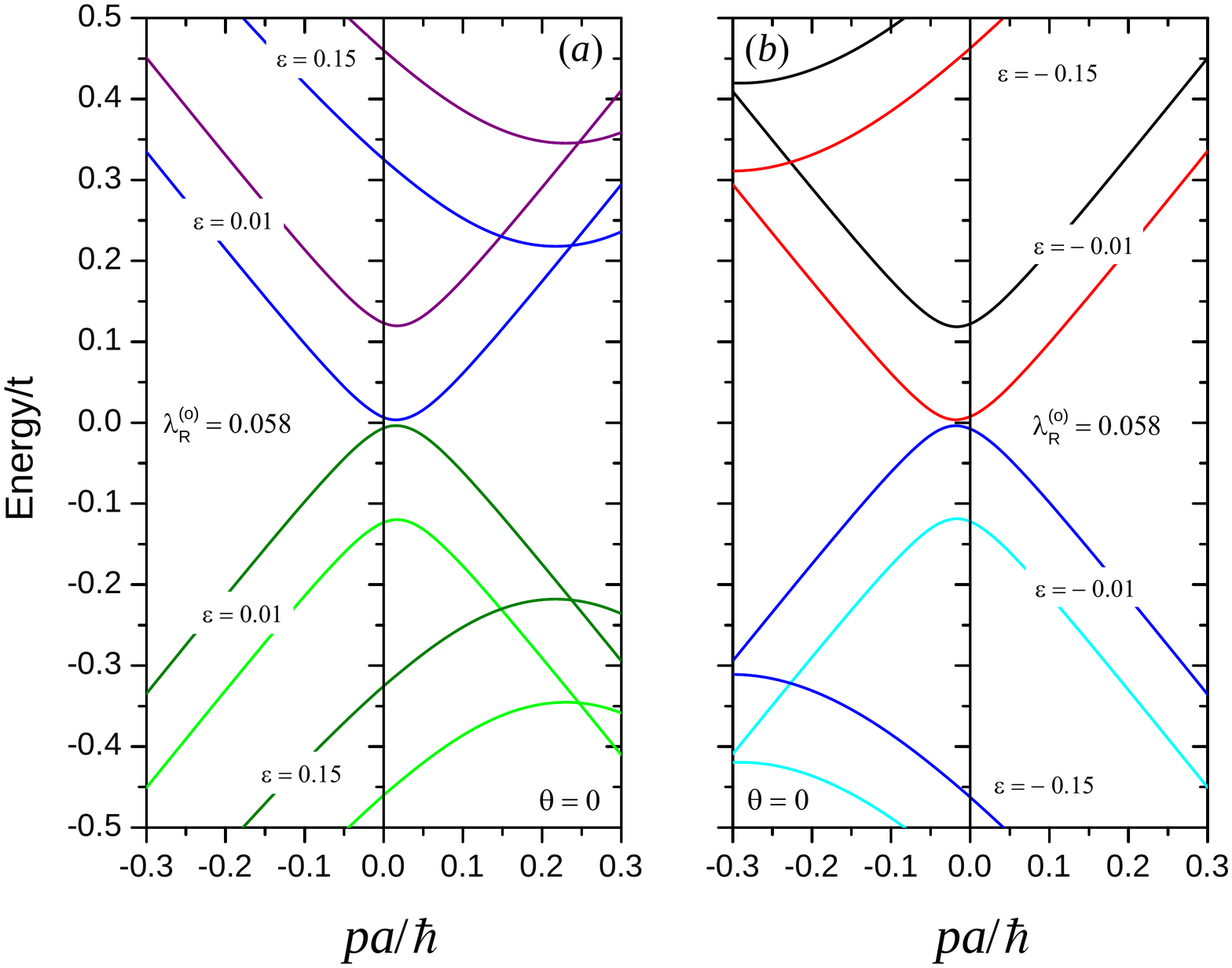}
\vspace{-0.1cm}
\caption{(Color online) Band structure for a graphene sheet under a simultaneous presence of relatively large Rashba spin-orbit coupling [$\lambda_{\text R}^{(0)}=0.0578$ in  units of $t$] and different values of uniaxial strains in the plane. The figure clearly shows how the chiral spin splitting increases
with the amplitude of the strain applied.}
\label{fig6}
\includegraphics[width=11cm]{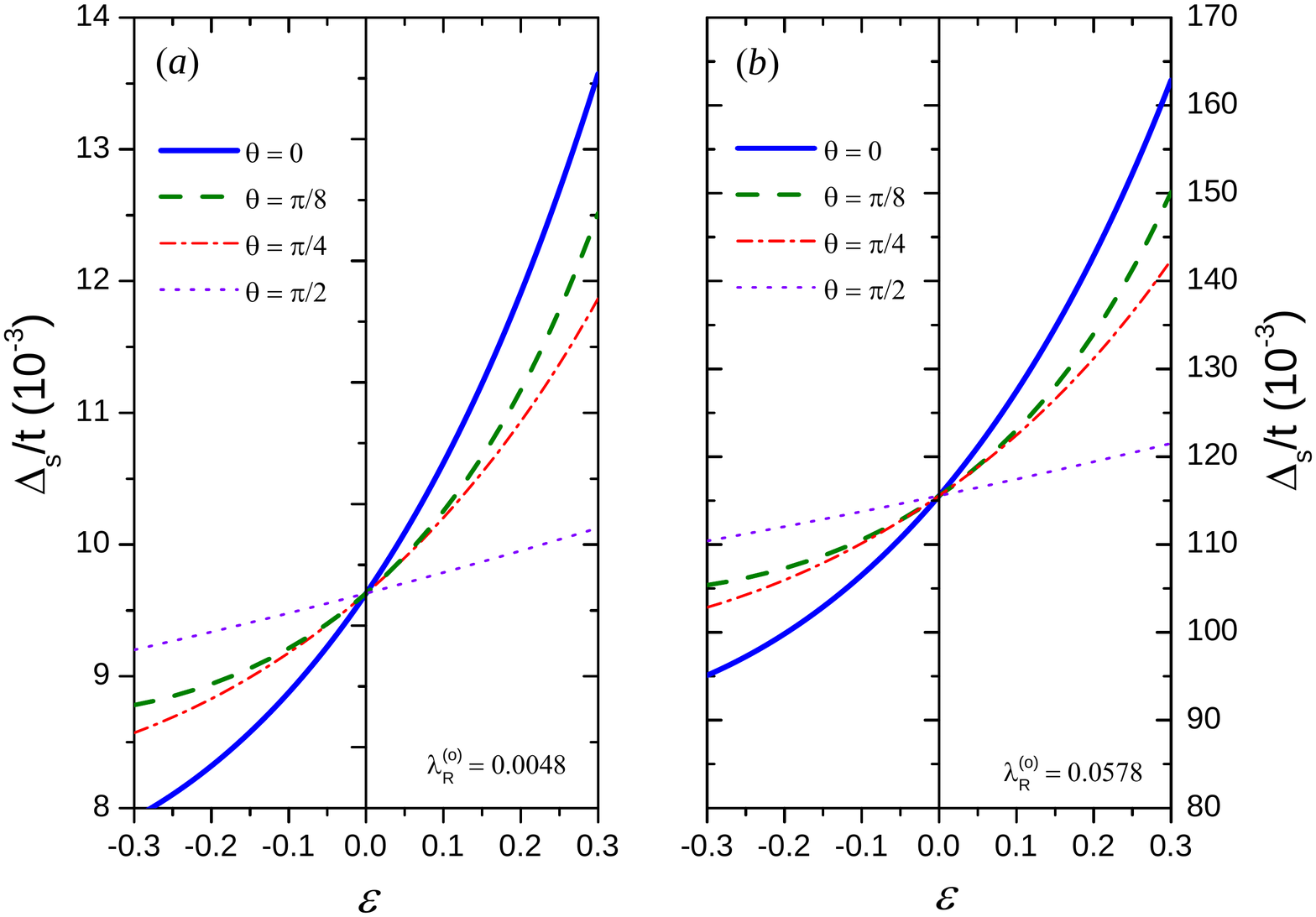}
\vspace{-0.1cm}
\caption{(Color online) Spin-splitting of the conduction band at $k=0$ for different orientation angles $\theta$. (a) For weak Rashba SOI $\lambda_{\text R}^{(0)}=0.0048$. (b) Large Rashba SOI $\lambda_{\text R}^{(0)}=0.058$. }
\label{fig7}
\end{center}
\end{figure}

In figure~\ref{fig4} we show the characteristic band dispersion as a function of $pa$ (with $p=p_x=p_y$) for different deformation strengths and for  Rashba coupling $\lambda_{\text R}^{0}=0.0048t$ and $\theta=0$. The plot clearly shows that even for a rather small strain there is a shifting of the band structure. It  varies with the orientation angle of the tension applied, not shown here. The presence of the strain opens a relatively large gap between the electron and hole bands and it can be significantly tuned by the strain. A similar behavior is shown in figure~\ref{fig5} and figure~\ref{fig6} where a larger Rashba SOI [$\lambda_{\text R}^{(0)}=0.024t$ and $\lambda_{\text R}^{(0)}=0.0578t$] has been used and where the gaps are consequently larger. We also observe that the gaps between $E_{+}^{-}$ and $E_{-}^{+}$ electron/hole bands can be continuously modulated from a semimetal to a semiconductor behavior just by varying the direction of the strains.

The dependence of the  spin-splitting energy $\Delta_s (k=0)=E_{+}^{+}(0)-E_{+}^{-}(0)$ of the conduction band structure at $k=0$  with the uniaxial strain $\varepsilon$  is shown in figure~\ref{fig7} for different orientation angle $\theta$  and (a) weak Rashba strength $\lambda_{\text R}^{(0)}=0.0048t$ and (b) strong Rashba strength $\lambda_{\text R}^{(0)}=0.0578t$.
Changes of the spin-splitting
can vary from 17\% to 59\% in the range of strains between $\varepsilon=-0.3$ and $\varepsilon=0.3$ studied here.

\section{Summary and conclusions}
We have developed a model to describe the spectral effects of small in plane deformations on gra\-phene,
in the vicinity of the K points. We focus on the non-intuitive enhancement of the Rashba SOI due to the stretching
of the sigma $s$-$p$ overlaps between the A--B sublattices. The deformations considered involve both bond
length and bond angle changes. The full Hamiltonian, to lowest order in the lattice momentum and
deformation amplitude, taking into account both kinetic energy and spin-orbit coupling
effects of deformations, can be cast into a convenient $\rm{U(1)}\times SU(2)$ non-Abelian gauge formulation.
We derive the analytical form for the energies as a function of the lattice momentum in the vicinity
of the K points and its dependence on bond stretching amplitude and angle.
Within reasonable lattice deformation strengths of at most 15\%, we find that the electron/hole bands
can be continuously modulated from a semimetal to a semiconductor, involving both shifts in the
position and size of the band gap. We also observe how the chiral spin splitting energy can be
controlled by changing bonds lengths and bond angles. The results found present an interesting prospect
for spintronics application for graphene on deformable substrates or charge induced lattice stretching
controlled by the gate voltages \cite{Wu14,Zhu15}.

\section*{Acknowledgement}
We are very pleased to have an opportunity to contribute to this special issue of Condensed Matter Physics in honor of our excellent friend Yurko (Yurij Holovatch).
EM and BB are respectively grateful to the University of Lorraine and to IVIC for invitations. They also thank the CNRS and FONACIT
for support through the ``PICS'' programme {\it Spin transport and spin manipulations in condensed matter: polarization, spin currents and entanglement}. BB and FM thank the DIONICOS
programme (FP7 IRSES) for financial support and FM acknowledges the project PAPIIT-UNAM IN111317.

\ukrainianpart

\title{Спін-орбітальна взаємодія Рашби, підсилена графеновими площинними деформаціями}
\author{Б. Берш\refaddr{label1,label2}, Ф. Мірелес\refaddr{label3}, E. Медіна\refaddr{label1,label2,label4}}
\addresses{
\addr{label1} Група статистичної фізики, Інститут Жана Лямура, Університет Лотарингії,\\ 54506 м. Вандувр-лє-Нансі, Франція
\addr{label2} Центр фізики, Інститут наукових досладжень Венесуели, 21827, м. Каракас, 1020 A, Венесуела
\addr{label3} Центр нанонаук і нанотехнологій, Національний автономний університет Мексики,  м. Мехіко, Мексика
\addr{label4} Ячай техноцентр, Вища школа фізичних наук і технології, Еквадор
}

\makeukrtitle

\begin{abstract}
Графен --- це моношаровий  вуглецевий кристал, де   2$p_z$ електрони демонструють лінійний закон дисперсії поблизу  рівня Фермі, що
описується безмасовим рівнянням Дірака у $2+1$ просторі-часі. Спін-орбітальні ефекти відкривають щілину в зонній структурі і вказують
на перспективи для керування спіном електронів провідності. Способи керування спін-орбітальним зв'язком в графені взагальному визначаються
близькістю ефектів до металів, які не поступаються мобільністю незбуреної системи та ймовірно індукують напруження в шарі графену.
В цій роботі ми досліджуємо $\rm{U(1)}\times SU(2)$ калібрувальні поля, які виникають з однорідного розтягнення графенового листа
під дією перпендикулярно-направленого електричного поля. Розгляд таких деформацій є особливо важливим через контрінтуїтивне підсилення зв'язку Рашби
в діапазоні 30-50\% для малих деформацій зв'язків, що є добре відомим з обчислень  в наближенні сильного зв'язку і з теорії функціоналу
густини. Ми повідомляємо досяжні зміни, які можуть бути здійснені в зонній структурі поблизу  K точок, як функцію  сили і
напрямку деформації.

\keywords{графен, спін-струм, спін-орбітальна взаємодія}
\end{abstract}


\begin{thebibliography}{50}


\bibitem{Peralta16} Peralta M., Colmenarez L., L\'opez A., Berche B., Medina E., Phys. Rev. B, 2016, {\bf 94}, 235407;\\ \bibdoi{10.1103/PhysRevB.94.235407}.

\bibitem{Pastawski} Calvo H.L., Pastawski H.M., Roche S., Foa Torres L.E.F., Appl. Phys. Lett., 2011, {\bf 98}, 232103; \bibdoi{10.1063/1.3597412}.

\bibitem{Zhou} Zhou S., Gweon G.-H., Lanzara A., Ann. Phys., 2006, {\bf 321}, 1730; \bibdoi{10.1016/j.aop.2006.04.011}.

\bibitem{KaneMele} Kane C.L., Mele E.J., Phys. Rev. Lett., 2005, {\bf 95}, 226801; \bibdoi{10.1103/PhysRevLett.95.226801}.

\bibitem{Wang07} Wang X.-F., Chakraborty T., Phys. Rev. B, 2007, {\bf 75}, 033408; \bibdoi{10.1103/PhysRevB.75.033408}.

\bibitem{Fertig} Brey L., Fertig H., Phys. Rev. B, 2006, {\bf 73}, 235411; \bibdoi{10.1103/PhysRevB.73.235411}.

\bibitem{Wang} Wang Z., Tang C., Sachs R., Barlas Y., Shi J., Phys. Rev. Lett., 2015, {\bf 114}, 016603;\\ \bibdoi{10.1103/PhysRevLett.114.016603}.

\bibitem{Macdonald} Chen H., Niu Q., Zhang Z., MacDonald A., Phys. Rev. B, 2013, {\bf 87}, 144410; \bibdoi{10.1103/PhysRevB.87.144410}.

\bibitem{MirelesEtAl} Carrillo-Bastos R., Faria D., Latg\'e A., Mireles F., Sandler N., Phys. Rev. B, 2014, {\bf 90}, 041411(R); \\ \bibdoi{10.1103/PhysRevB.90.041411}.

\bibitem{AndoSO} Ando T., J. Phys. Soc. Jpn., 2000, {\bf 69}, 1757; \bibdoi{10.1143/JPSJ.69.1757}.

\bibitem{Marchenko} Marchenko D., Varykhalov A., Scholz M.R., Bihlmayer G.,
Rashba E.I., Rybkin A., Shikin A.M., Rader O., Nat. Commun., 2012, {\bf 3}, 1232; \bibdoi{10.1038/ncomms2227}.

\bibitem{Huertas} Huertas-Hernando D., Guinea F., Brataas A., Phys. Rev. B, 2006, {\bf 74}, 155426; \bibdoi{10.1103/PhysRevB.74.155426}.

\bibitem{Konschuh} Konschuh S., Gmitra M., Fabian J., Phys. Rev. B, 2010, {\bf 82}, 245412; \bibdoi{10.1103/PhysRevB.82.245412}.

\bibitem{JinLiZhang06} Jin P.Q., Li Y.Q., Zhang F.C., J. Phys. A: Math. Gen., 2006, {\bf 39}, 7115; \bibdoi{10.1088/0305-4470/39/22/022}.

\bibitem{MedinaLopezBerche08} Medina E., L\'opez A., Berche B., EPL, 2008, {\bf 83}, 47005; \bibdoi{10.1209/0295-5075/83/47005}.

\bibitem{Tokatly08} Tokatly I.V.,  Phys. Rev. Lett., 2008, {\bf 101}, 106601; \bibdoi{10.1103/PhysRevLett.101.106601}.

\bibitem{BercheChatelainMedina} Berche B., Chatelain C., Medina E., Eur. J. Phys., 2010, \textbf{31}, 1267; \bibdoi{10.1088/0143-0807/31/5/026}.

\bibitem{BercheMedinaEJP} Berche B., Medina E., Eur. J. Phys., 2013, {\bf 34}, 161; \bibdoi{10.1088/0143-0807/34/1/161}.

\bibitem{WuEtAl2016} Wu Q-P., Liu Z.-F., Chen A.-X., Xiao X.-B., Liu Z.-M., Sci. Rep., 2016, {\bf 6}, 21590; \bibdoi{10.1038/srep21590}.

\bibitem{PastawskiMedina} Pastawski H.M., Medina E., Rev. Mex. Fis., 2001, {\bf 47}, 1.

\bibitem{VarelaEtAl16} Varela S., Mujica V., Medina E., Phys. Rev. B, 2016, {\bf 93}, 155436; \bibdoi{10.1103/PhysRevB.93.155436}.

\bibitem{Vozmediano} Vozmediano M.A.H., Katsnelson M.I., Guinea F.,  Phys. Rep., 2010, {\bf 496}, 109; \bibdoi{10.1016/j.physrep.2010.07.003}.

\bibitem{CastroNetoArxiv} Pereira V.M., Castro Neto A.H., Preprint \arxiv{0810.4539v3}, 2009.

\bibitem{AndoPhonons} Suzuura H., Ando T., Phys. Rev. B, 2002, {\bf 65}, 235412; \bibdoi{10.1103/PhysRevB.65.235412}.

\bibitem{KaneMele2} Kane C.L., Mele E.J., Phys. Rev. Lett., 2005, {\bf 95}, 146802; \bibdoi{10.1103/PhysRevLett.95.146802}.

\bibitem{Varykhalov} Varykhalov A., Sanchez-Barriga J., Shikin A.M., Biswas C., Vescovo E., Rybkin A., Marshenko D., Rader O., Phys. Rev. Lett., 2008, {\bf 101}, 157601; \bibdoi{10.1103/PhysRevLett.101.157601}.

\bibitem{Dedkov} Dedkov Y.S., Fonin M., R\"udiger U., Laubschat C., Phys. Rev. Lett., 2008, {\bf 100}, 107602;\\ \bibdoi{10.1103/PhysRevLett.100.107602}.

\bibitem{Rader} Rader O., Varykhalov A., Sanchez-Barriga J., Marshenko D.,  Rybkin A., Shikin A.M.,  Phys. Rev. Lett., 2009, {\bf 102}, 057602; \bibdoi{10.1103/PhysRevLett.102.057602}.

\bibitem{Pereira-Neto} Pereira V.G., Castro Neto A.H., Peres N.M.R., Phys. Rev. B, 2009, {\bf 80}, 045401; \bibdoi{10.1103/PhysRevB.80.045401}.

\bibitem{Poisson_ratio} Landau L.D., Lifshitz E.M., Theory of Elasticity, 3rd Edn., Reed Educational and Professional Publishing Ltd., Oxford, 1986.

\bibitem{Blakslee} Blakslee L., Seldin E.J., Stence G.B., Wen T., J. Appl. Phys., 1970, {\bf 41}, 3373; \bibdoi{10.1063/1.1659428}.

\bibitem{CastroNetoGuinea07} Castro Neto A.H., Guinea F., Phys. Rev. B, 2007, {\bf 75}, 045404; \bibdoi{10.1103/PhysRevB.75.045404}.

\bibitem{YangMills54} Yang C.N., Mills R.L., Phys. Rev., 1954, {\bf 96}, 191; \bibdoi{10.1103/PhysRev.96.191}.

\bibitem{Wu14} Wu Z., Huang X.-L., Wang Z.-L., Xu J.-J., Wang H.-G., Zhang X.-B., Sci. Rep., 2014, {\bf 4}, 3669; \bibdoi{10.1038/srep03669}.

\bibitem{Zhu15} Zhu S., Stroscio J.A., Li T., Phys. Rev. Lett., 2015, {\bf 115}, 245501; \bibdoi{10.1103/PhysRevLett.115.245501}.

\end{thebibliography}
\end{document}